\documentclass[a4paper,11pt]{article}
\pdfoutput=1 

\usepackage{jinstpub} 

\title{ Design of Sc-ECAL Prototype for CEPC and Performance of First Two Layers}

\author[a,b]{Yazhou Niu,}
\author[a,b]{Shensen Zhao,}
\author[a,b,1]{Yunlong Zhang,\note{Corresponding author.}}
\author[a,b]{Zhongtao Shen}
\author[a,c]{Mingyi Dong,}
\author[a,c]{Zhigang Wang,}
\author[a,b]{Yukun Shi,}
\author[a,b]{Anshun Zhou }
\author[d]{Zhen Wang}
\author[a,b]{Jianbei Liu}
\author[a,b]{Shubin Liu}

\affiliation[a]{State Key Laboratory of Particle Detection and Electronics, China}
\affiliation[b]{Department of Modern Physics , University of Science and Technology of China, HeFei 230026, China}
\affiliation[c]{Institute of High Energy Physics, Chinese Academy of Sciences (CAS), Beijing 100049, China}
\affiliation[d]{Shanghai Jiao Tong University, Shanghai 200240, China}

\emailAdd{ylzhang@ustc.edu.cn}

\abstract{As a future Higgs or Z factory, construction of the circular electron positron collider (CEPC) has been proposed to precisely measure Higgs bosons. A particle flow-oriented electromagnetic calorimeter (ECAL) and a hadronic calorimeter (HCAL) comprise a baseline design for the CEPC calorimetry system. The scintillator-tungsten ECAL (Sc-ECAL) prototype is being developed within the CEPC calorimetry working group. The Sc-ECAL is a sampling calorimeter consisting of alternating absorber layers and high granularity active layers. Each component of the active layer has been studied and optimized from a variety of aspects to meet the design requirements. The complete technological Sc-ECAL prototype will contain 30 layers of ECAL-based unit (EBU) boards alternating with 30 layers of the absorber. The first two layers of the prototype have been developed, and primary commissioning has been performed to validate their performance.}


\arxivnumber{ 2002.01809 } 

\collaboration[c]{}

\proceeding{3$^{\text{th}}$ Workshop on CHEF2019\\
  25-29 November 2019\\
  Kyushu University, Japan}

\begin{document}
\maketitle
\flushbottom

\section{Introduction}
\label{sec:intro}

\paragraph{} The circular electron positron collider (CEPC) has been proposed as a Higgs or Z factory that would be able to conduct precision measurements on the Higgs and  Z bosons~\cite{1}. The potential physics programs of the CEPC~\cite{2} require all possible final states, which are decays from intermediate vector bosons, to be reconstructed with unprecedented precision. In particular, the requirement for a jet energy resolution of $\sim$30\%/$\sqrt{E}$ at an energy below 100 GeV is a primary motivation for the development of the particle flow algorithms (PFA)~\cite{3,4} oriented for high granularity calorimeters. For an electromagnetic calorimeter (ECAL), the energy resolution is required to be better than  $\sim$16\%/$\sqrt{E}$ for photons, and the unit cell size no more than 10 mm for the shower separation ability requirement~\cite{5}.

Over the past a dozen years, the CALICE Collaboration have been working on the PFA-orientd ECAL and HCAL in different areas of technology to develop high performance detectors for high energy e$^{+}$e$^{-}$ experiments . The scintillator-tungsten electromagnetic calorimeter (Sc-ECAL), one choice of PFA-oriented ECAL technologies, is the baseline design of ECAL in CEPC CDR~\cite{1}. For the present, Sc-ECAL with front-end electrtonic readout is still not studied yet on full-scale prototype. One full-scale technological prototype is essential to build to research the Sc-ECAL technology and performance.


\section{Design of Sc-ECAL}

\begin{figure}[htbp]
\centering 
\includegraphics[width=.8\textwidth,clip]{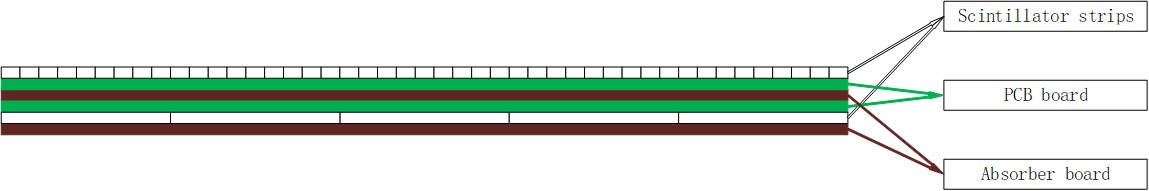}
\qquad
\includegraphics[width=.4\textwidth]{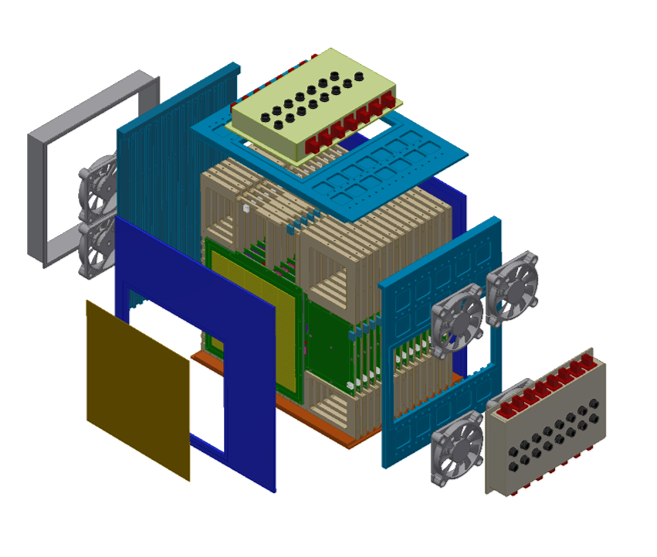}
\caption{\label{fig:prototype} The layout of sandwich structure and exploded view of the full-scale Sc-ECAL prototype}
\end{figure}

\paragraph{} Sc-ECAL is a PFA-oriented sampling calorimeter with a high granularity cell size. Sensitive layer alternating with absorber layer constitutes the sandwich structure. Detector units integrated with electronic board form one ECAL-based unit (EBU) board. The full technological ECAL prototype consists of 15 super-layers. Each super-layer contains 2 layers EBU and the corresponding data interface (DIF) board, inserted one tungsten-copper alloy absorber board located in the middle of the sandwich structure. Thus, there are 30 EBU boards and 30 DIF boards in total. One EBU sensitive detector comprises 210 scintillator strips coupled with a silicon photomultiplier (SiPM). SiPM readout by fornd-end SPIROC2E~\cite{6} chip, which integrates 36 channels. Eventually, there will be 6300 channels read by 180 SPIROC2E chips.  Figure~\ref{fig:prototype} presents the layout of the sandwich structure (top), and a diagram of the exploded view of the Sc-ECAL prototype (bottom).

\subsection{Scintillator strip}
\paragraph{}The shower separation ability is crucial for the performance of a PFA-oriented ECAL, which can reconstruct the jet energy resolution with unprecedented precision. Based on $\nu$$\nu$Higgs$\rightarrow$gluons studies, ECAL cell size is recommended to be equal to or smaller than 10 mm~\cite{5}. The Sc-ECAL scintillator strip dimensions are 45 mm $\times$ 5 mm $\times$ 2 mm, and it is arranged crosswise on adjacent layers, such that a 5 mm $\times$ 5 mm cell size is formed. A new bottom-center coupling mode~\cite{7} has been proposed to eliminate the non-uniformity, especially close to the side-end. The dimensions of the dimple have been iteratively designed and optimized according to the measurement and simulation light output performances. Finally, the light output non-uniformity can achieve 4\% with 17\% variance for different strips. For $\nu$$\nu$Higgs$\rightarrow$$\gamma$$\gamma$, the reconstructed Higgs mass resolution degrades by only 2.5\% in comparison with perfect uniformity~\cite{8}.

\subsection{SiPM}
\label{sec:SiPMs}
\paragraph{} One of the greatest challenges for the Sc-ECAL is the dynamic range demand up to 800 minimum ionized particle (MIPs). SiPM is a new type of photon sensor that has been widely used in particle physics experiments~\cite{9}. The SiPM can achieve saturation with the number of incident photons increases, which allows the SiPM to detect photons more than the number of pixels through a correction. The typical light output of one scintillator strip is ~20 photon-electrons per MIP~\cite{7}, such that the SiPM have to potentially detect 16,000 photon-electrons. The larger number of pixels an SiPM has, the larger dynamic range can be obtained. Two types SiPM with 10,000 and 4,500 pixels, produced by Hamamatsu~\cite{10}, have been adopted by Sc-ECAL technological prototype. The pixel size of two types SiPM are  10-$\upmu$m and  15-$\upmu$m respectively. The later type SiPM have more fewer pixels number but more higher gain comparing with the first one. Fewer pixels SiPM could be placed on before shower start and the shower tail, where the demand of dynamic range is not that high.
\subsection{Electronics}
\paragraph{} Embedding electronics is the only method to meet the critical space requirements for CEPC ECAL. The SPIROC~\cite{6} chip is a dedicated front-end application specific integrated circuit (ASIC) for the SiPM readout, which developed by the OMEGA group. The SPIROC chip has been design to meet the demand of large dynamic range SiPM readout.The chip have developed from the first version SPIROC1 to the second iteration which implements all features and corrects some bugs, and the subversion SPIROC2E which corrects the bugs observed of the SPIROC2B. Each SPIROC chip has 36 channels and each channel has 16 buffer memory cells. A low power consumption of 8 mW per channel and a large dynamic range of 62 fC--300 pC can meet the demand of Sc-ECAL units' dynamic range. The test board, both based on single SPIROC2B chip and 4-chips, have been developed and studied in 2018 ~\cite{11}. For the SPIROC2E chip, from the beginning of 2019, development evolved from a single chip to a 6-chip single layer board. Finally, readout electronics technological board with 6 SPIROC2E chips has been prototyped. The preliminary performance of this board will be presented in section~\ref{sec:EBUperformance}.


\section{ Performance of the First Two Layers }
\begin{figure}[htbp]
\centering 
\includegraphics[width=.4\textwidth,trim=0 0 0 0,clip]{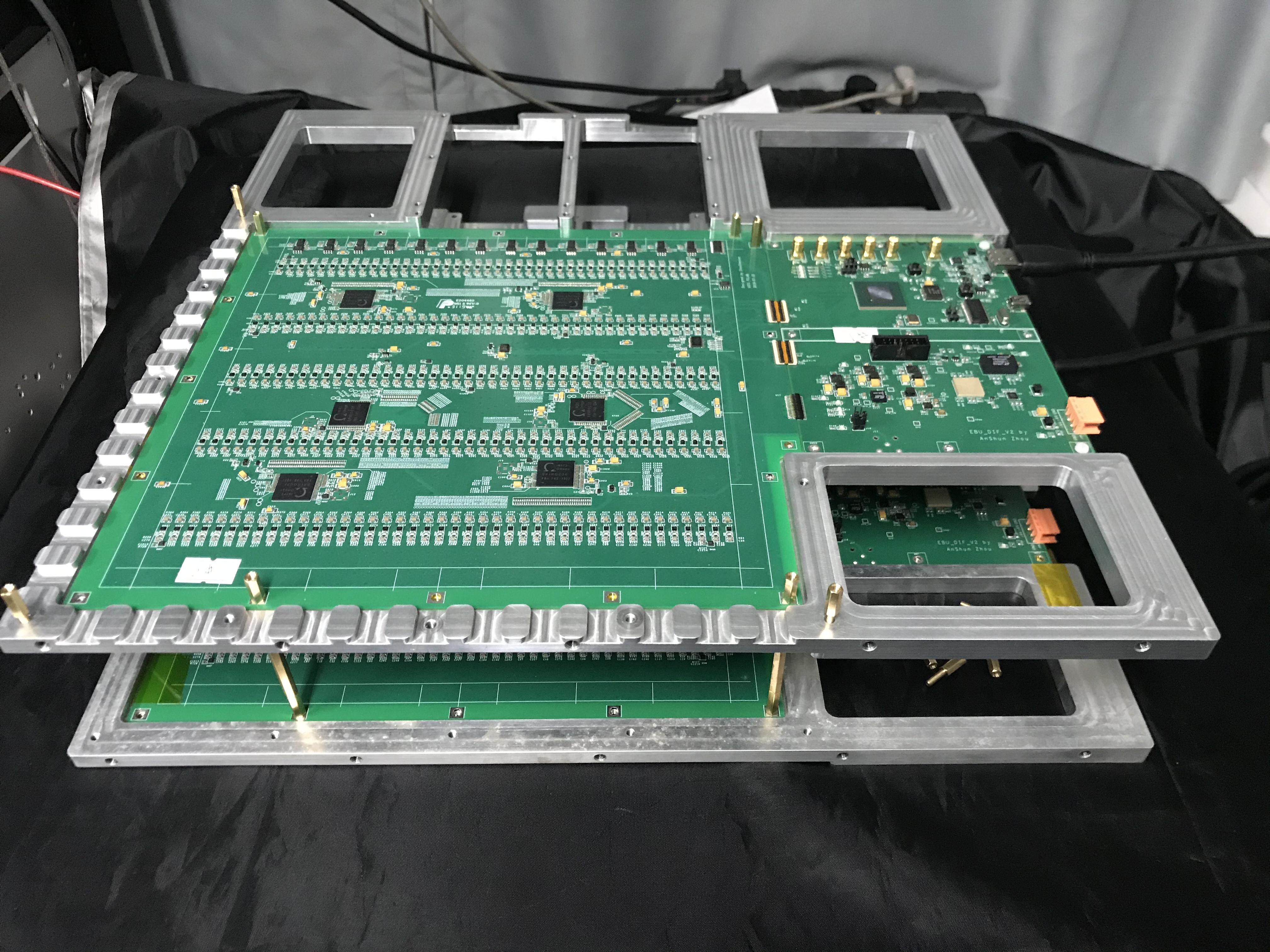}
\qquad
\includegraphics[width=.4\textwidth,trim=0 0 0 0,clip]{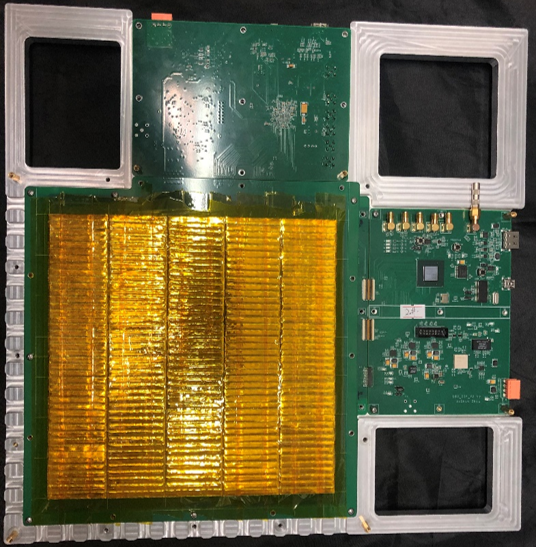}
\caption{\label{fig:superlayer} Electronics side view of the EBU (left) and the detector side view of the integrated super-layer (right) }
\end{figure}

\paragraph{} Currently, as figure~\ref{fig:superlayer} presents, two layers of EBU have been produced and assembled into one superlayer. The total thickness of one layer EBU is controlled to be below 6 mm within a 1-mm tolerance, excluding the DIF board. There are two independent calibration systems, charge inject calibration and LED calibration, and both work properly. A temperature monitoring system and SiPM operation voltage compensation are also available in this EBU and behave well. The first two layers of the EBU have two types of SiPM respectively, as described in section~\ref{sec:SiPMs}. This paper will present the EBU electronics performance and MIP measurement results by $\beta$ radiation source in the following subsections.

\subsection{Pedestal}
\label{sec:EBUperformance}
\begin{figure}[htbp]
\centering 
\includegraphics[width=.4\textwidth,trim=0 0 0 0,clip]{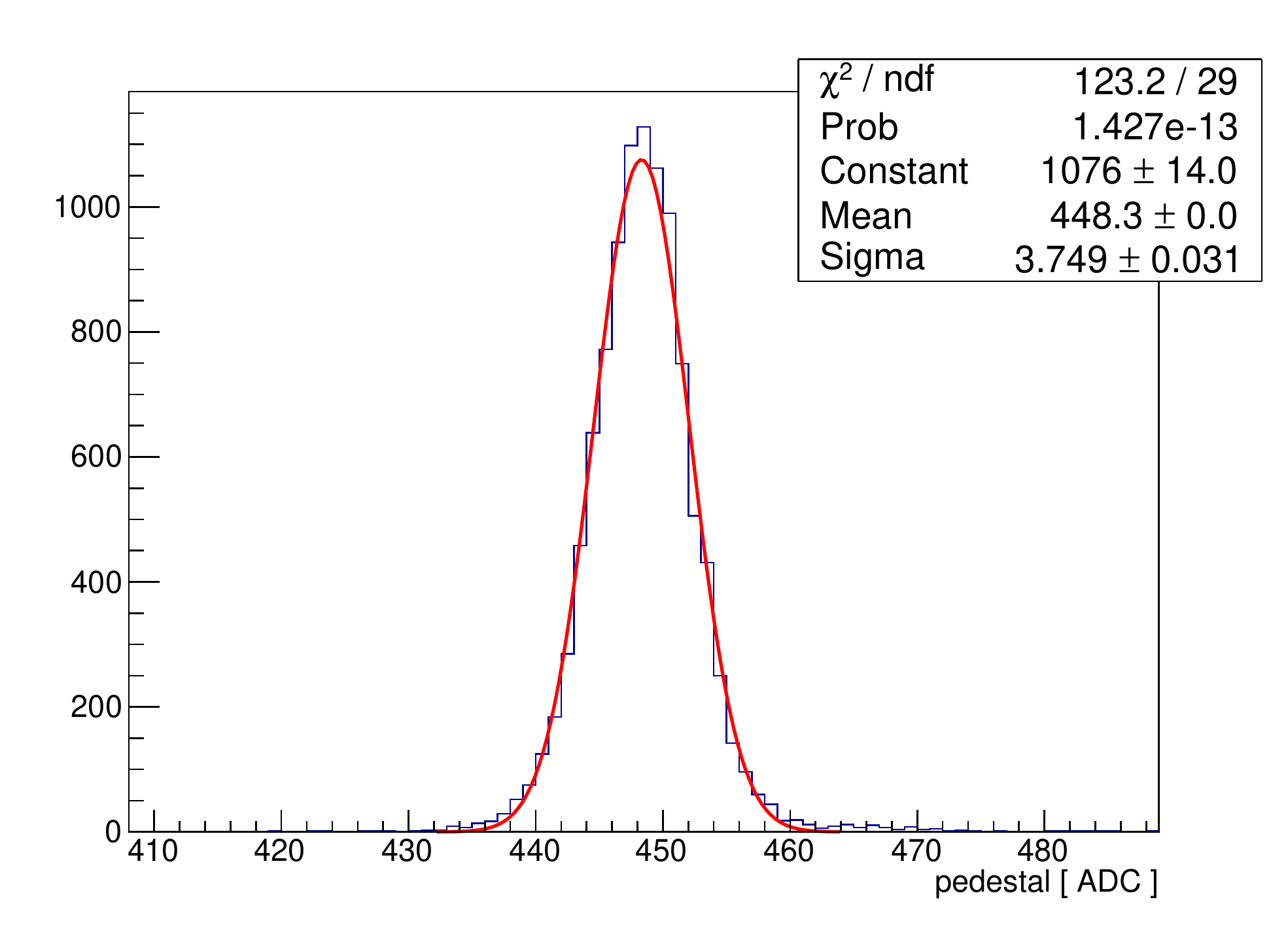}
\qquad
\includegraphics[width=.4\textwidth,trim=0 0 0 0,clip]{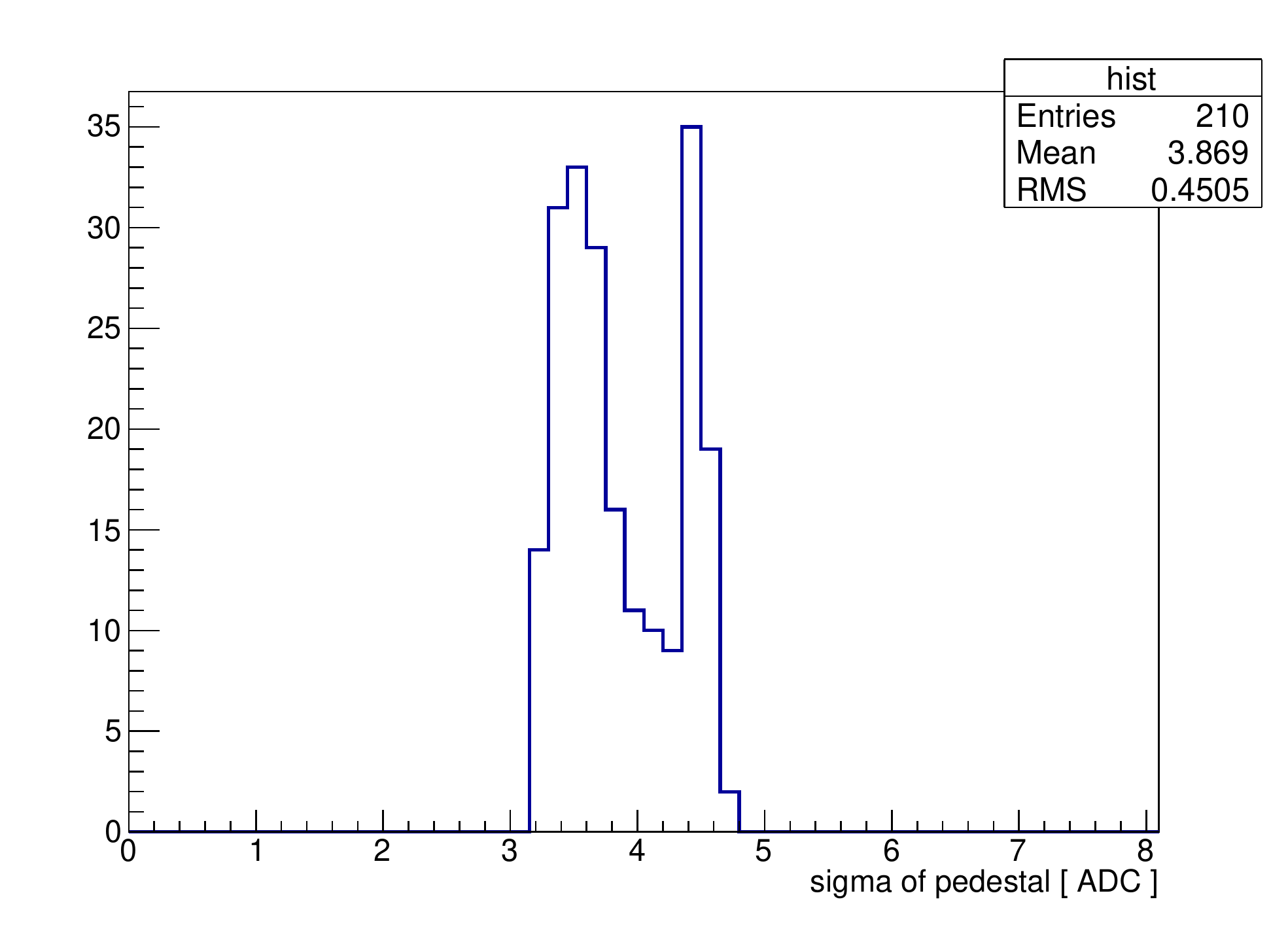}
\caption{\label{fig:pedestal}Sigma for the electronics pedestal (left) and distribution for one EBU (right).}
\end{figure}

\paragraph{}
The left portion of figure ~\ref{fig:pedestal} gives an example of one channel pedestal spectrum fit by a Gause function, and the right shows the 210 channels per one EBU layer. The typical sigma of the pedestal is about 3.8 ADC, with about 10\% variance for one EBU.

\begin{figure}[htbp]
\centering 
\includegraphics[width=.45\textwidth,trim=0 0 0 0,clip]{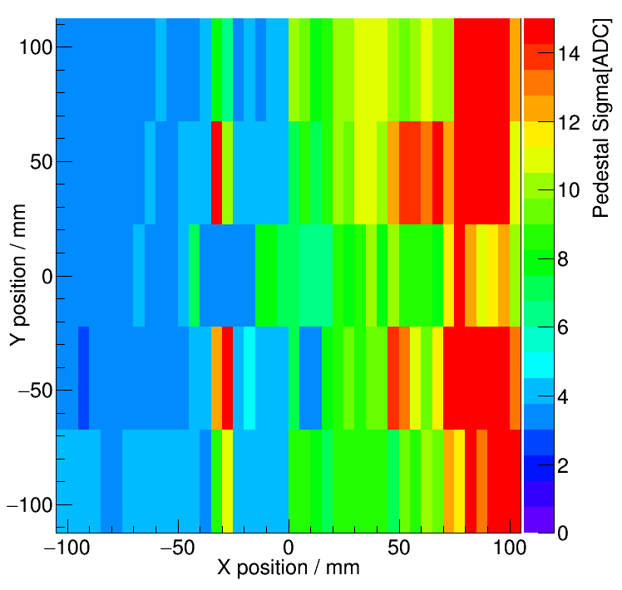}
\qquad
\includegraphics[width=.45\textwidth,trim=0 0 0 0,clip]{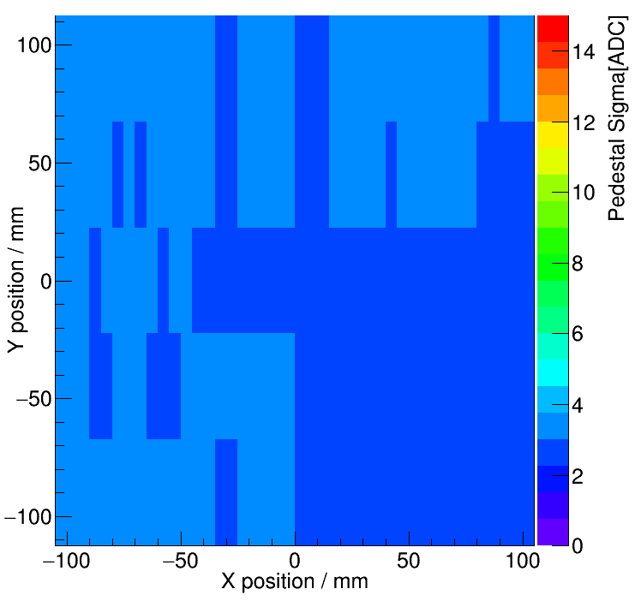}
\caption{\label{fig:compensation}Map of the sigma for the pedestal without (left) and with operational voltage compensation (right). }
\end{figure}

The width of the pedestal for different channels has a significant influence on the signal measurement during a test.The left map of figure ~\ref{fig:compensation} illustrates the sigma of the pedestal when the SiPM works at a uniform voltage, and the map on the right shows the results after compensation. Through comparison with and without a SiPM working voltage compensation, it is evident that the width of the electronics pedestal is strongly dependent on the working voltage of the SiPM. Thus, the SPIROC2E compensation pin's voltage must be adjusted to enable each SiPM to work on its operational voltage, i.e., the same overvoltage. After compensation, the sigma of the pedestal has no numerical value and no chip dependence.

\subsection{MIP results}
\begin{figure}[htbp]
\centering 
\includegraphics[width=.4\textwidth,trim=0 0 0 0,clip]{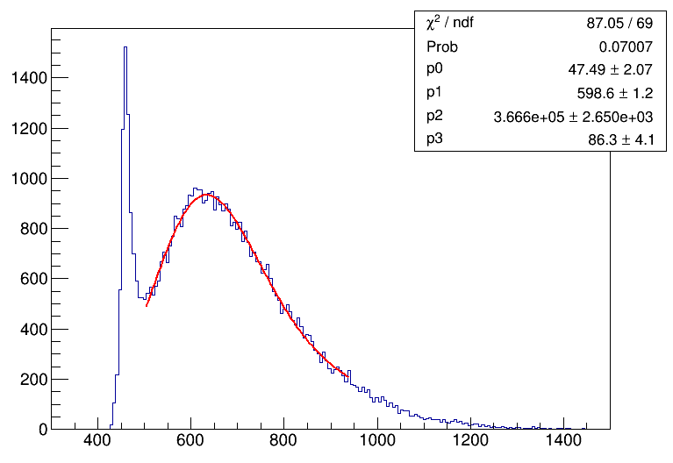}
\qquad
\includegraphics[width=.4\textwidth,trim=0 0 0 0,clip]{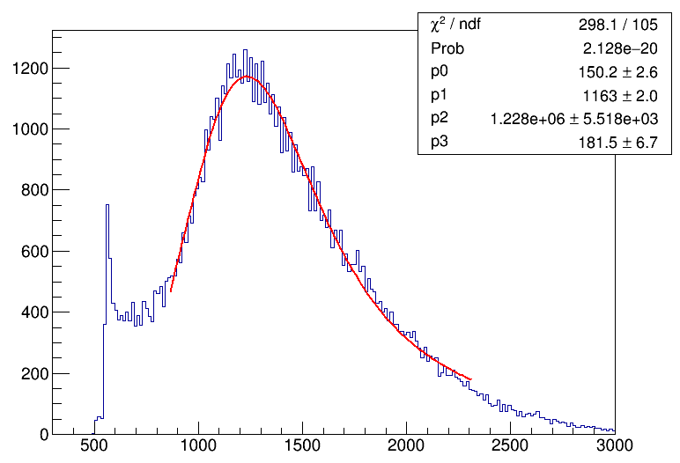}
\caption{\label{fig:MIP} Typical MIP signal for 10 $\upmu$m pixel size SiPM (left) and 15 $\upmu$m pixel size SiPM (right) before pedestal substraction.}
\end{figure}

\paragraph{} The parameter set points are identical for both EBUs, namely, 25 ns shaping time and an auto-trigger with 16 memory cells. Measurement with a $^{90}$Sr radiation source, as Figure~\ref{fig:MIP} presents, shows a typical MIP signal distribution fit by a Landau convolution with Gauss. The value of subtracting the pedestal from the most possible value (MPV) of the fitting resultes provides the MIP response. 

\begin{figure}[htbp]
\centering 
\includegraphics[width=.4\textwidth,trim=0 0 0 0,clip]{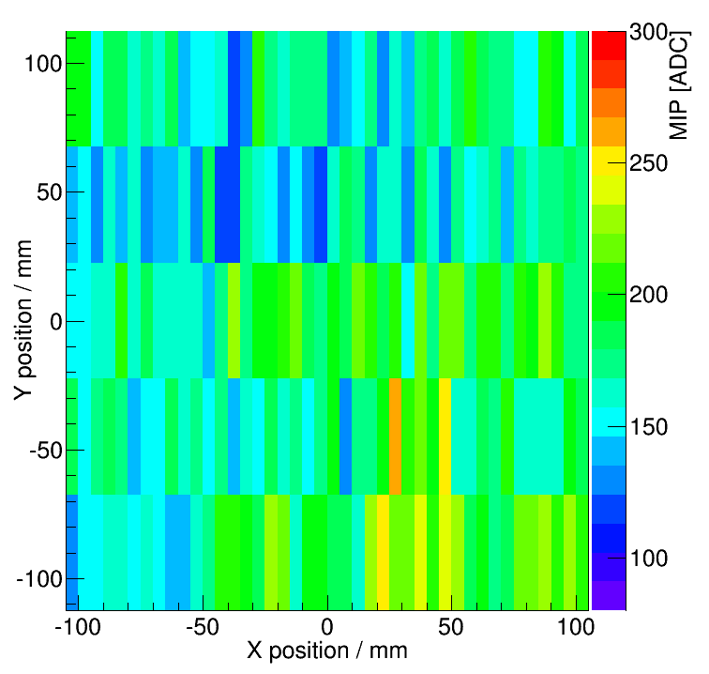}
\qquad
\includegraphics[width=.4\textwidth,trim=0 0 0 0,clip]{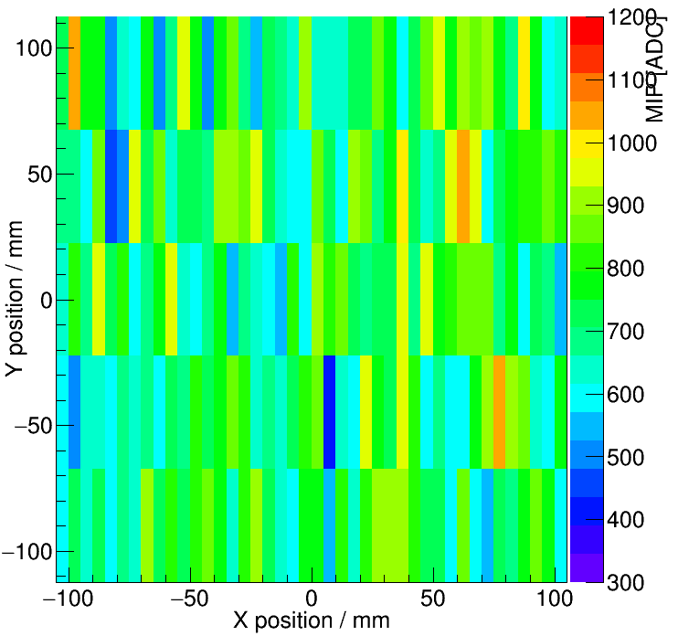}
\caption{\label{fig:MPV} MIP signal MPV for 10-$\upmu$m EBU (left) and 15-$\upmu$m EBU (right).}
\end{figure}

Scanning for all channels, the MIP response and uniformity of the first two EBU layers have been obtained. As Figure~\ref{fig:MPV} shows, the MIP signal of two layers EBU 420 channels well separate. Additionally, the MIP response distribution with position indicates that the result is position independent. The variation for all MIP response channels is about 16\% for both EBUs. The signal-over-noise ratio, i.e., MIP MPV over the sigma of the pedestal, is about 35 for a 10-$\upmu$m pixel-size SiPM EBU and 135 for a 15-$\upmu$m pixel-size SiPM EBU in the channel level. Two main reasons for the difference are the gain of the 15-$\upmu$m being about 1.7 times the 10-$\upmu$m SiPM and the photon detect efficiency of the 15-$\upmu$m SiPM being about 2.5 times the 10-$\upmu$m SiPM. Thus, the MIP response for the 15-$\upmu$m SiPM is about 4.2 times the MIP response of the 10-$\upmu$m SiPM.

%
%

\section{Summary}

\paragraph{} Sc-ECAL is a PFA-oriented sampling calorimeter with a high granularity cell size. Front-end SPIROC2E chip readout SiPM which collects the light output of the scintillator strips. The design of Sc-ECAL for each component have been done and the performance have been studied through simulation. The first two layers of the prototype have been developed and commissioning to validate the performance of the integrated EBU. From the preliminary results of the first two layers of the prototype, 100\% of the channels well separate the MIP signal. The MIP MPV variation is about 16\% and has no position dependence. The performance achieves the desired aims, the complete 30-layer Sc-ECAL prototype is being developed and will be performance-tested in 2020 to further validate its performance.


\acknowledgments

The author wishes to thank all colleagues in the CEPC Calorimetry Group for their excellent cooperation and help in the making of the prototype. We appreciate Shanghai Institute of Ceramic for assembling the scintillator strips on EBU, especially Junfeng Chen and Xiaojian Song. This study was supported by the National Key Program for S\&T Research and Development (Grant No.: 2016YFA0400400) and National Natural Science Foundation of China (11635007).



\end{document}